\documentclass[twoside,a4paper]{article}
\usepackage{fleqn,epsf}
\usepackage{authblk}
\usepackage{graphicx}
\usepackage{subfigure}
\usepackage{amsmath}
\usepackage{cite}
\usepackage[usenames]{color}
\usepackage{lineno,hyperref}
\usepackage{soul}
\usepackage[section]{placeins}
\usepackage{float}

\begin{document}

\title{Hit time and hit position reconstruction in the J-PET detector based on a library of averaged model signals}

\author{P.~Moskal$^a$, N.~G.~Sharma$^a$, M.~Silarski$^a$, T.~Bednarski$^a$, P.~Bia{\l}as$^a$, J. Bu{\l}ka$^b$ ,E.~Czerwi{\'n}ski$^a$,
A.~Gajos$^a$, D.~Kami{\'n}ska$^a$ , L.~Kap{\l}on$^{a,c}$, A.~Kochanowski$^d$, G.~Korcyl$^a$, J.~Kowal$^a$,
 P.~Kowalski$^e$, T.~Kozik$^a$, W.~Krzemie{\'n}$^a$, E. Kubicz$^a$, Sz.~Nied{\'z}wiecki$^a$, M. Pa{\l}ka$^a$,
L.~Raczy{\'n}ski$^e$, Z.~Rudy$^a$, O.~Rundel$^a$, P.~Salabura$^a$, A.~S{\l}omski$^a$, J.~Smyrski$^a$, A.~Strzelecki$^a$,
A.~Wieczorek$^{a,c}$, W.~Wi{\'s}licki$^e$, I. Wochlik$^b$ ,M.~Zieli{\'n}ski$^a$, N.~Zo{\'n}$^a$
 }
\affil{$^a$Faculty of Physics, Astronomy and Applied Computer Science, Jagiellonian University,
30-348 Cracow, Poland\\
$^b$Department of Automatics and Bioengineering AGH, Poland\\
$^c$Institute of Metallurgy and Materials Science of Polish Academy of Sciences,
30-059 Cracow, Poland.\\
$^d$Faculty of Chemistry, Jagiellonian University, 30-060 Cracow, Poland\\
$^e${\'S}wierk Computing Centre, National Centre for Nuclear Research,
05-400 Otwock-\'Swierk, Poland\\

}
\maketitle
\begin{abstract}
In this article we present a novel method of hit time and hit position reconstruction in long scintillator detectors. We
take advantage of the fact that for this kind of detectors amplitude and shape of registered signals depends strongly on
the position where particle hit the detector. The reconstruction is based on determination of the degree of similarity
between measured and averaged signals stored in a library for a set of well-defined positions along the scintillator.
Preliminary results of validation of the introduced method with experimental data obtained by means of the double strip
prototype of the J-PET detector are presented.
\end{abstract}
{PACS:87.57.uk, 07.05.Kf}
%keywords: Positron Emission Tomography, Time-of-flight, reconstruction, J-PET

\section{Introduction}
% The rapid development of new methods in Positron Emission Tomography
% results also in a progress in many other fields, in particular in development of
% new hit time and position reconstruction.
Recently a new concept of large acceptance Jagiellonian PET (J-PET) system was proposed~\cite{Pm1,Pm2,Pm3,Pm4}. Unlike all
the commercial PET devices using inorganic scintillators as radiation detectors~\cite{M,J,Js,D} (usually these are the BGO,
LSO or LYSO crystals) J-PET is based on the polymer scintillators. This technique offers improvement of the Time of Flight
(TOF) resolution\footnote{ Presently best TOF resolution was achieved with LSO crystals and amounts to about
400~ps~\cite{W}.} and also constitutes a promising solution for a whole-body PET imaging. In the case of J-PET annihilation
gamma quanta are registered by means of long scintillator strips read out from both ends by photomultipliers. This allows
for the determination of position and time of the gamma quanta reaction based predominantly on the time measurement.
Therefore, to fully exploit the potential of this solution, it requires the elaboration of a new hit position and TOF
reconstruction methods~\cite{lech}. Recently, one method of reconstruction in scintillator detectors based on the
comparison of registered signals with respect to a library of synchronized model signals collected for a set of
well-defined positions along the scintillator~\cite{Nat} is published. In this article we present similar method, however
the comparison of measured signal is done with averaged model signals determined as a function of position along the
scintillator. This approach speeds up significantly the reconstruction in comparison to the previously used method.
\section{Library of synchronized model signals}
\label{exp}
\begin{figure} [h]
\begin{center}
 \includegraphics*[width=0.7\textwidth]{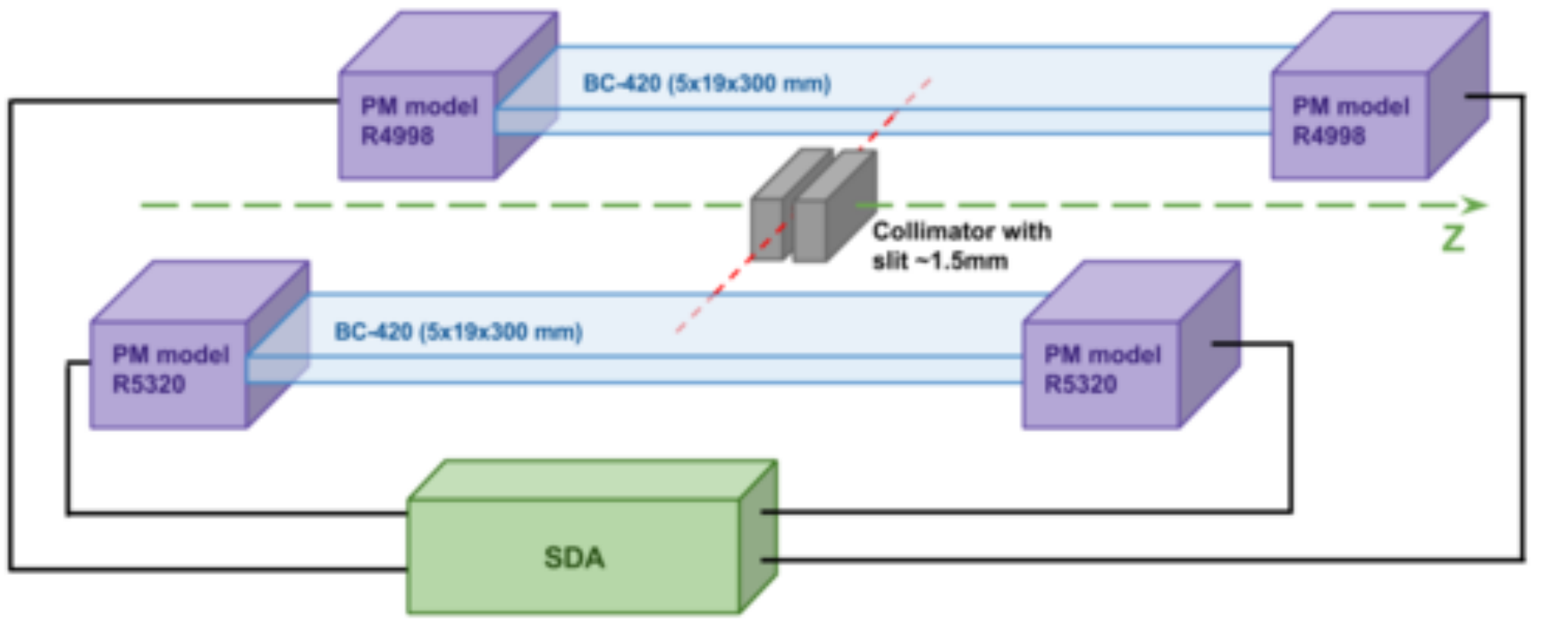}
 \caption{Scheme of experimental setup where abbrevations PM and SDA
denote Photomultilplier and Serial Data analyser (SDA6000A), respectively.} \label{Fig1}
\end{center}
%\label{Fig1}                        help to float cross reference if one put label just after figure caption.
\end{figure}

In order to create the library of model signals, we have used the setup composed of two BC-420~\cite{Saint}
scintilators with dimensions 300 x 19 x 5 mm$^3$ and Hamamatsu photomultipliers R4998~\cite{Hamamatsu} connected optically
to their most distant ends via optical gel EJ-550~\cite{scionix}. General scheme of experimental setup used to build the
library of signals is shown in Fig.~\ref{Fig1}.

%General scheme of
%experimental setup used to build the library of signals is shown in Fig.~\ref{Fig1}. The setup is composed of two
%BC-420~\cite{Saint} scintilators with dimensions 300 x 19 x 5 mm$^3$ and Hamamatsu photomultipliers R4998~\cite{Hamamatsu}
%connected optically to their most distant ends via optical gel EJ-550~\cite{scionix}.
%
\begin{figure*}[h]
           \centerline{
           \subfigure[!htb][]
           {
              \includegraphics*[width=0.5\textwidth]{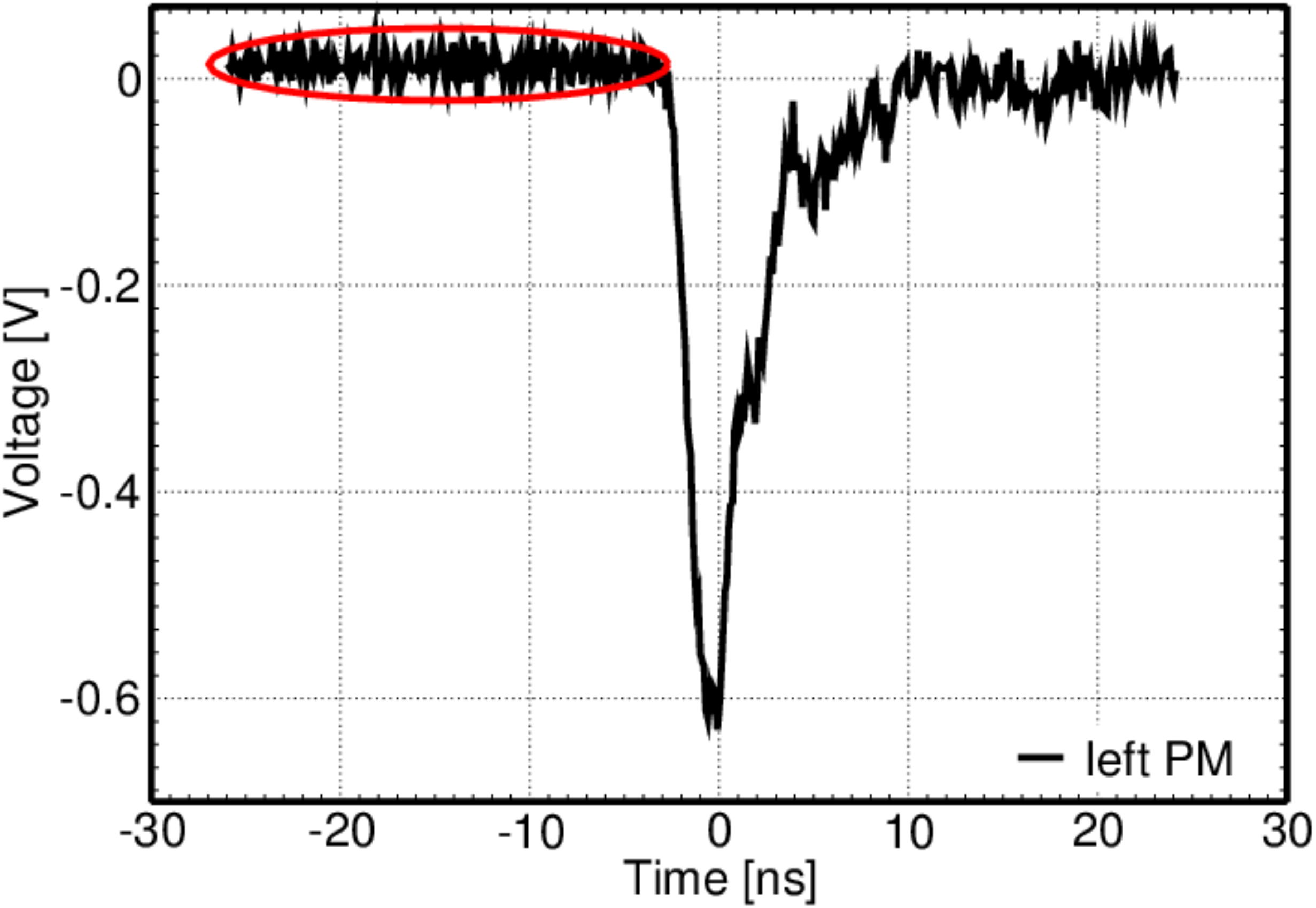}
               \label{F(a)}
           }
           \subfigure[!htb][]
           {
               \includegraphics*[width=0.5\textwidth]{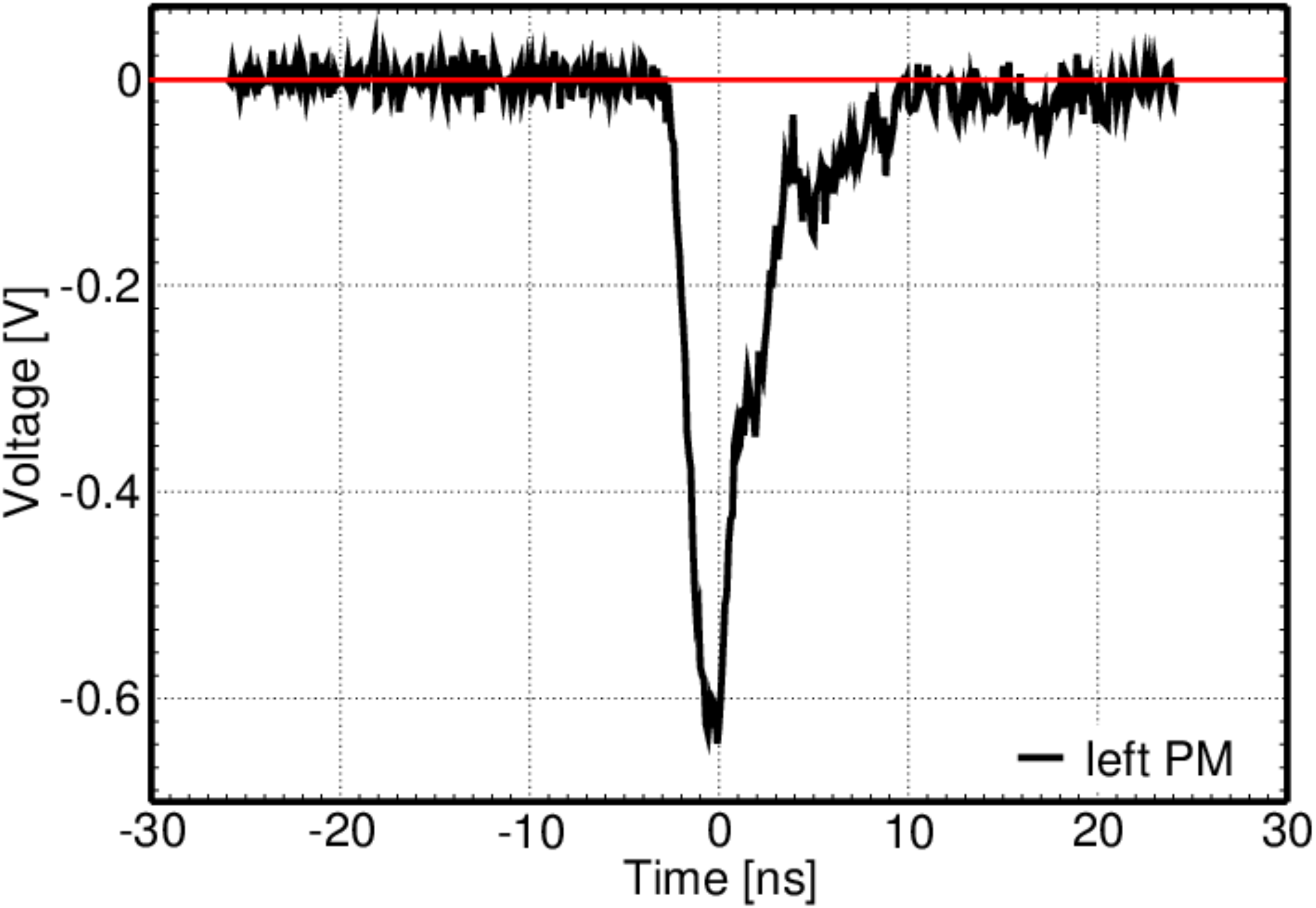}}
                \label{F(b)}
           }
              \caption{(a) Measured signal from the database
  (b) The same signal after pedestal correction.}
\label{Fig2}
\end{figure*}

\begin{figure}[h]
\begin{center}
 \includegraphics*[scale=0.30]{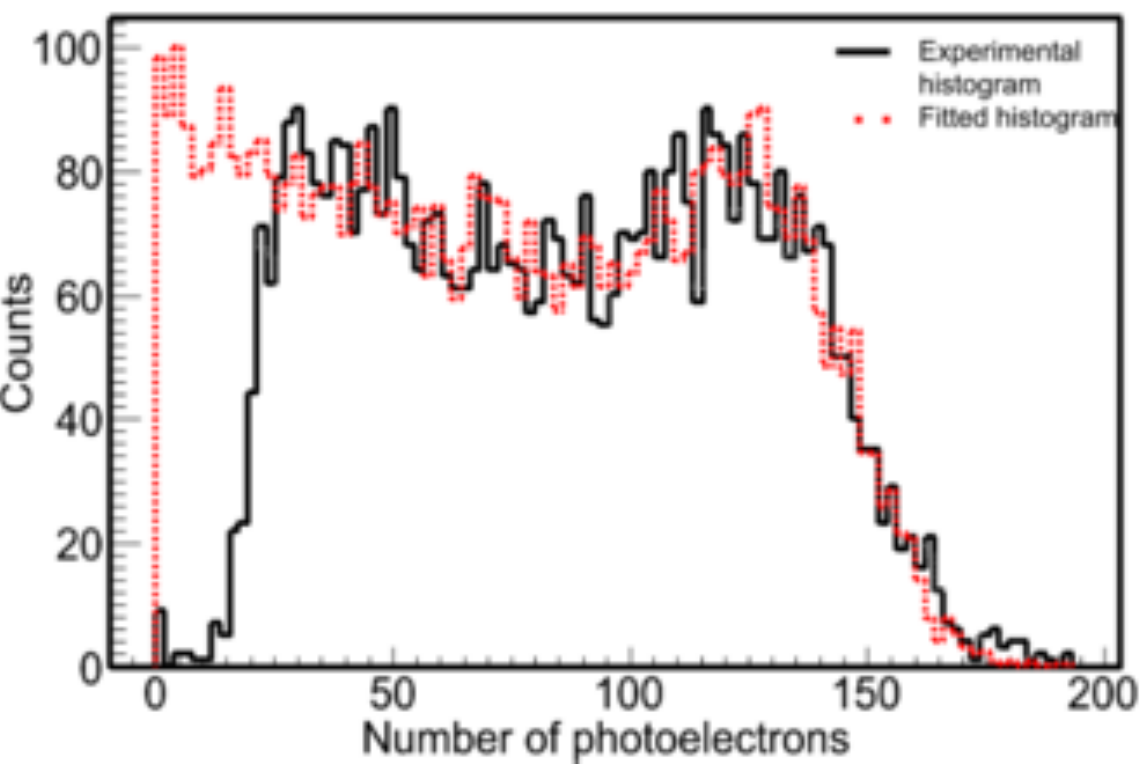}
 \caption{Distribution of number of photoelectrons obtained from signals measured irradiating
the scintillator strip at central position(solid black histogram). Dashed red histogram represents theoretical
distribution~\cite{O} fitted to the experimental data using normalization and energy calibration constants as free
parameters. Details of the fitting procedure are described in reference~\cite{Pm6}.}\label{Fig3}
\end{center}
%\label{Fig3}
\end{figure}

The two detectors were irradiated with a collimated beam of annihilation quanta (FWHM $\sim$~1.5 mm) in steps of 3 mm. The
$^{22}$Na source was placed inside a lead collimator which could be moved along the scintillator by using a dedicated
mechanical system. For each position a high statistics of signals with interval of 100 ps was collected by means of SDA. A
coincident registration of signals from both detectors allowed for the photomultipliers noise suppression and the selection
of annihilation gamma quanta.
%Signals registered by photomultipliers were probed with 100~ps intervals by means of Serial Data Analyzer (LeCroy SDA6000A).
Exemplary sampled signal is shown in Fig.~\ref{Fig2}a.
%
%Before creating the library few modifications were done in the measured database
%of signals in order to suppress background, and thus to construct a background-free library.
%To do this
To create a background-free library, first we have corrected all the signals for pedestal.
 For every signal the average value of voltage was calculated in the noise region
showed in Fig.~\ref{Fig2}a by encircled red area.
This computed average value was then used for pedestal correction for that particular signal.
An exemplary signal after correction is presented in Fig.~\ref{Fig2}b.
After pedestal correction, signals building the library were selected based on number of registered photoelectrons such
that it is constructed from events with the high energy deposition. In Fig.~\ref{Fig3} we show an example of such spectrum
obtained at central position. For every position we have cut all the signals for which the number of registered
photoelectrons is lower than half of the number of photoelectrons corresponding to the Compton edge for 511~keV gamma
quanta.
Finally, database signals were synchronized by shifting their time scales in such a way that time of the gamma quantum hit
inside the detector is the same for all events in the library.\\
The shape of a model signal for each position from the database is determined by averaging pedestal corrected and
synchronized signals.
%Such two average signals determined at both sides of scintillator are treated as a reference signal
%for the reconstruct position.
More details on model signals determination are given in the next section of this article.
\section{Determination of the shape of model signals}
\label{met} 
\begin{figure*}[h]
           \centerline{
           \subfigure[!htb][]
           {
              \includegraphics*[width=0.5\textwidth]{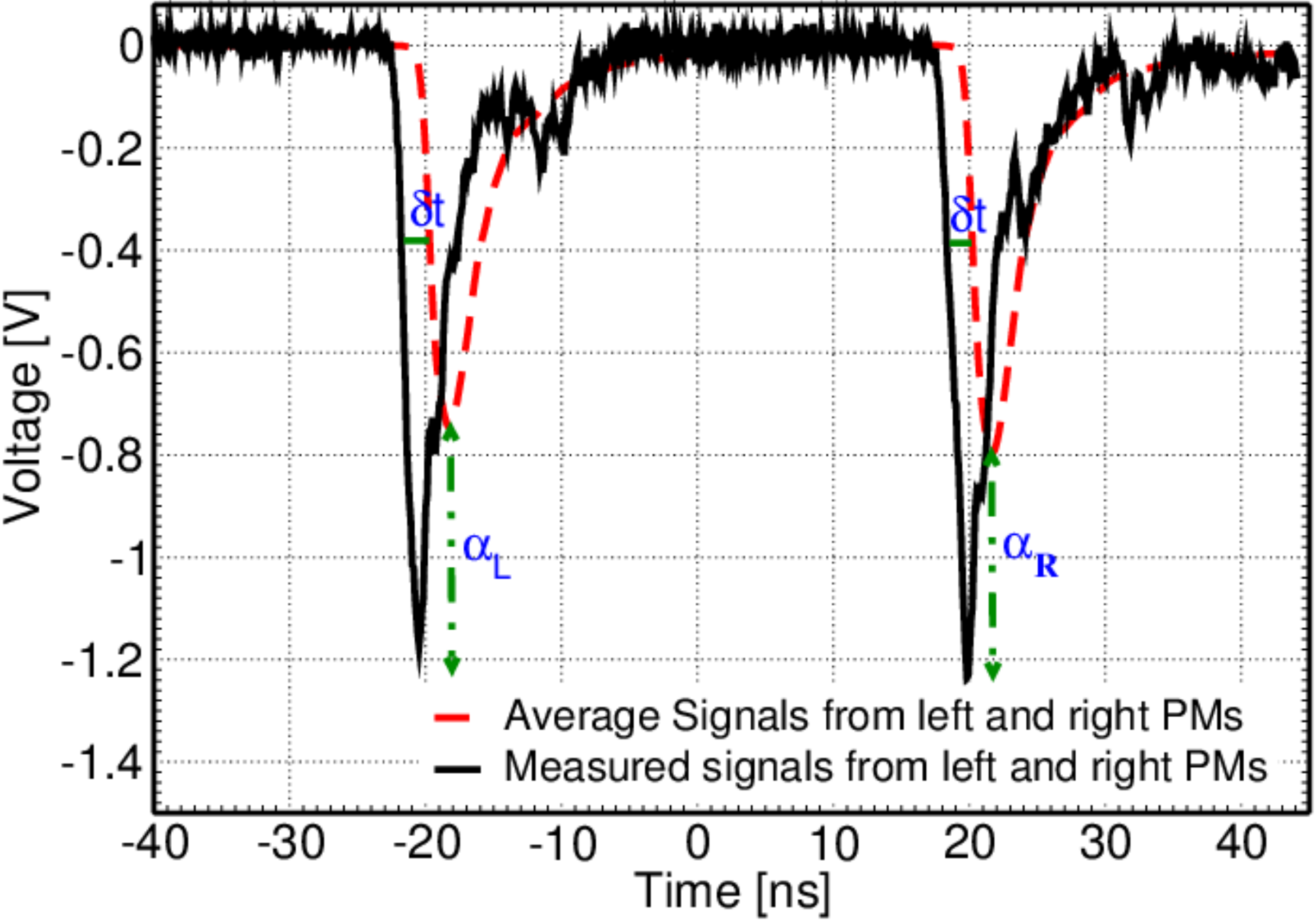}
               \label{Fig(a)}
           }
           \subfigure[!htb][]
           {
               \includegraphics*[width=0.5\textwidth]{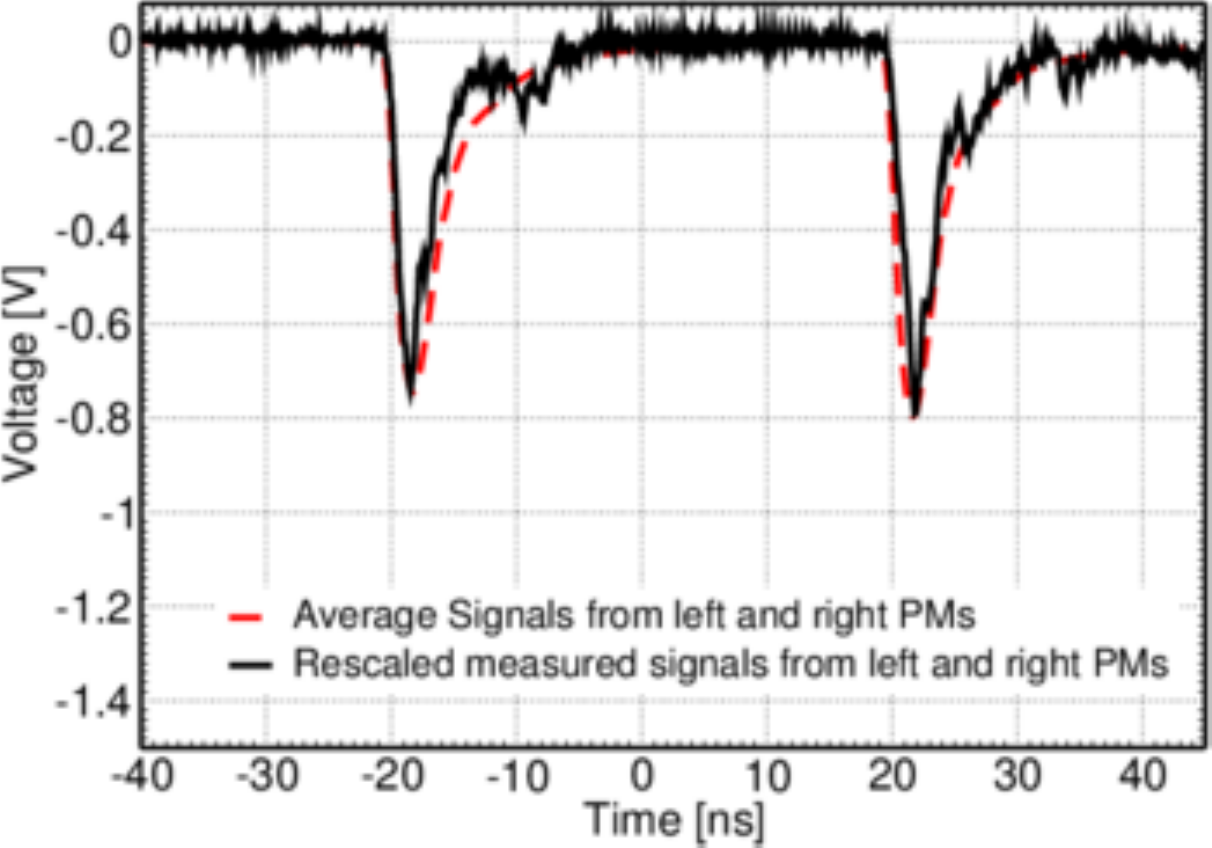}}
                \label{Fig(b)}
           }
              \caption{(a) Example of a database signal before alignment to the average. Black curve
                            represents the measured signal, while the average signal is shown by red dashed curve.
  (b) The same signal after rescaling.}
    \label{Fig4}
\end{figure*}

\begin{figure}[h]
 \center{}
\includegraphics[width=0.55\textwidth]{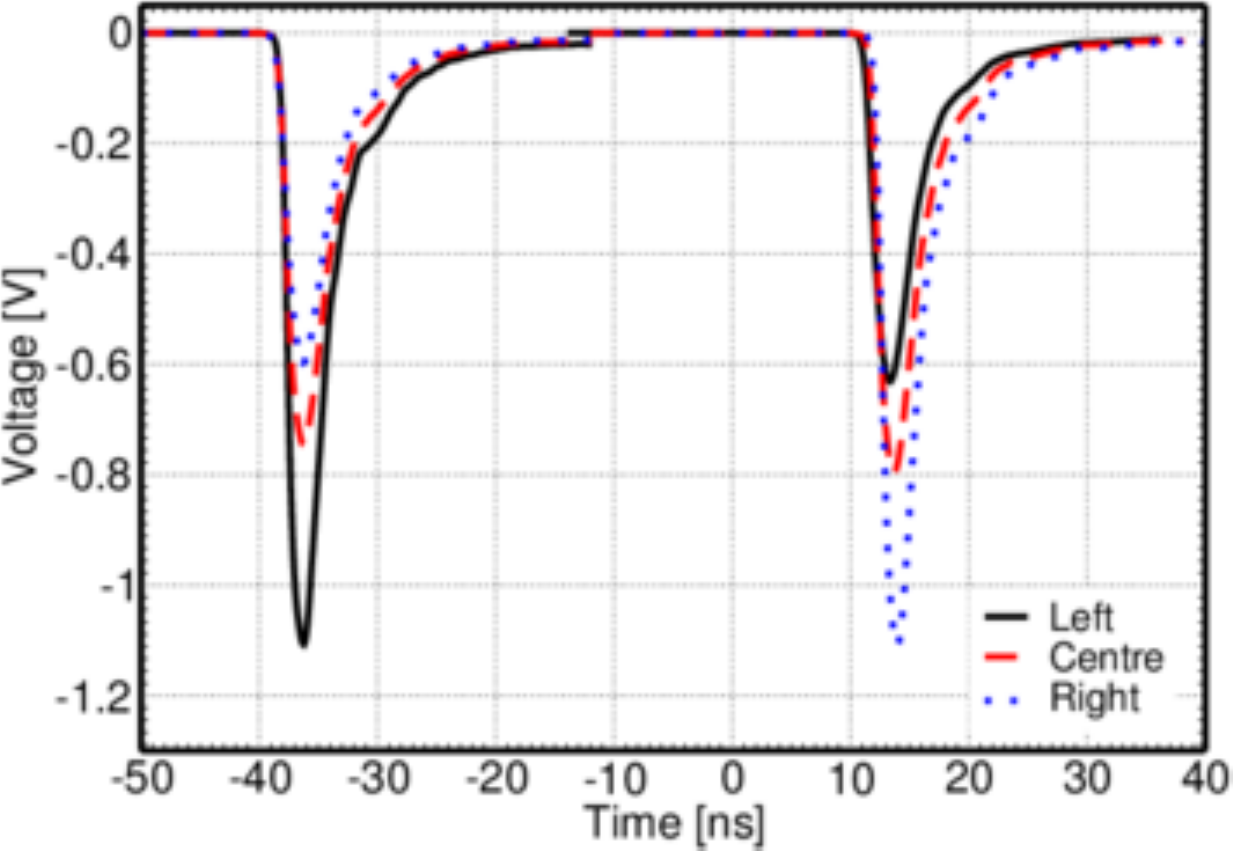}
\caption{Sample of model signals at three different positions: solid black line represents the left side of the
scintillator, dashed red line center and dotted blue line the right side of scintillator.} \label{Fig5}
\end{figure}

Function describing the approximate shape of the model signal was determined by averaging measured signals for
a given hit position. Calculated average for signals registered simultaneously at both ends of the scintillator was treated
as a reference in order to align all the database signals measured for a particular position, as it is shown in
Figs.~\ref{Fig4}a and b.
Such alignment is necessary to account for the spread of signals in amplitude and time. To perform signal's alignment in
the database, we have defined for each measured signal at a given position  a $\chi^2$ value. It was computed by comparing
leading edge of the database and average signals. Comparison was performed by taking into account signal registered at the
left and right side of scintillator simultaneously. The $\chi^2$ was calculated as a function of three parameters
$\delta$t, $\alpha_L$ and $\alpha_R$:
\begin{eqnarray}
%\begin{align}
\label{eqn2} \chi^2(\delta t, \alpha_L, \alpha_R) = \sum_{i=1}^{n}\frac{(t_{ModelLeft}(V_i) - t_{dbLeft}((\alpha_L V_i)
- \delta t ))^2}{n}\\
+ \sum_{i=1}^{m}\frac{(t_{ModelRight}(V_i) - t_{dbRight}((\alpha_R V_i) - \delta t ))^2}{m}~. \nonumber
%\end{align}
\end{eqnarray}
$\delta$t is a time shift for sample points along time axis and $\alpha_L$, $\alpha_R$ are normalisation factors for both
signals (left and right) registered at both ends of scintillator. $t_{ModelLeft}(V_i)$ and $t_{ModelRight}(V_i)$ denote
time for model signal computed at left and right side for voltage $V_i$ at their leading edge. $t_{dbLeft}((\alpha_L V_i)$
and $t_{dbRight}((\alpha_R V_i)$ is the time computed for rescaled left and right signals at their leading edge,
respectively. By minimization of the $\chi^2$ value we obtained the best alignment between two compared signals and each
database signal was rescaled using fit parameters $\alpha_L$, $\alpha_R$ and $\delta t$ giving the lowest $\chi^2$. Next,
average of these rescaled signals was computed again leading to the model signal determination and the whole procedure was
repeated until the changes were negligible. Example of model signals obtained at three different positions is shown in
Fig.~\ref{Fig5}.
\section{Validation and optimization of the time and position reconstruction}
\label{valid} In order to reconstruct hit position of gamma quanta in the detector the measured signals are compared with
each model signal from the library. This is done by minimizing $\chi^2$ value calculated at the leading edges of measured
and model signals. Validation and optimization of this reconstruction method has been performed by utilizing the signals
gathered at known positions with the same experimental setup which was used to build the library. In order to simulate the
response of front-end electronics we have determined the times corresponding to predefined constant level
thresholds~\cite{fpga}: -60 mV, -120 mV, -180 mV and -240 mV. Analogously, we have chosen also four constant fraction
thresholds to sample signals at: 1/8$A_i$, 2/8$A_i$, 3/8$A_i$, 4/8$A_i$, where $A_i$ is the amplitude of signal which is
smallest among all four compared signals. To optimize the reconstruction we have considered two posibilities:
\begin{enumerate}
  \item when $\chi^2$ is a function of the time shift $\delta t$ only:
  \begin{eqnarray}
    \label{eqn3}
%   \begin{align*}
     \chi^2(\delta t) = \sum_{i=1}^{4}(t_{ModelLeft}(V_i) - t_{dbLeft}(V_i -\delta t ))^2\\
        + \sum_{i=1}^{4}(t_{ModelRight}(V_i) - t_{dbRight}(V_i - \delta t ))^2
%     \end{align*}
    \nonumber
     \end{eqnarray}
  \item when $\chi^2$ depends on $\delta t$ and normalisation factors $\alpha_L$ and $\alpha_R$:
    \begin{eqnarray}
    \label{eqn4}
%   \begin{align*}
     \chi^2(\delta t, \alpha_L, \alpha_R) = \sum_{i=1}^{4}(t_{ModelLeft}(V_i) - t_{dbLeft}((\alpha_L V_i)
     - \delta t ))^2\\
      + \sum_{i=1}^{4}(t_{ModelRight}(V_i) - t_{dbRight}((\alpha_R V_i) - \delta t ))^2~.
%     \end{align*}
\nonumber
     \end{eqnarray}
 \end{enumerate}
$t_{ModelLeft}(V_i)$, $t_{ModelRight}(V_i)$ and $t_{dbLeft}(V_i)$, $t_{dbRight}(V_i)$ denote times determined at left and
right side of the scintillator at threshold $V_i$ for model and registered signals, respectively. The reconstructed hit
position is the position of most similar signal from the library with respect to measured signal(i.e. model signal for
which $\chi^2$ is minimal\footnote{
Since the degree of similarity is represented by the $\chi^2$ value.}).\\
The time of particle interaction is determined as a relative time between the measured signal and the most similar one from
the library. This provides also determination of the gamma quantum time of flight $(TOF)$ ~\cite{Nat}:
\begin{eqnarray}
%\begin{align*}
\nonumber
t_{firstStrip} = \delta t _{firstStrip}\\
t_{secondStrip} = \delta t _{secondStrip}\\
%\end{align*}
%\begin{equation}
\nonumber
\mathrm{TOF} = t_{secondStrip} - t_{firstStrip}~,
%\end{equation}
\end{eqnarray}
where $\delta t$ denotes shift in time for which the computed $\chi^2$ defined in Eq.~\ref{eqn3} and \ref{eqn4} is lowest.
\begin{figure}[h]
\begin{center}
%\centering
 \includegraphics[width=0.65\textwidth]{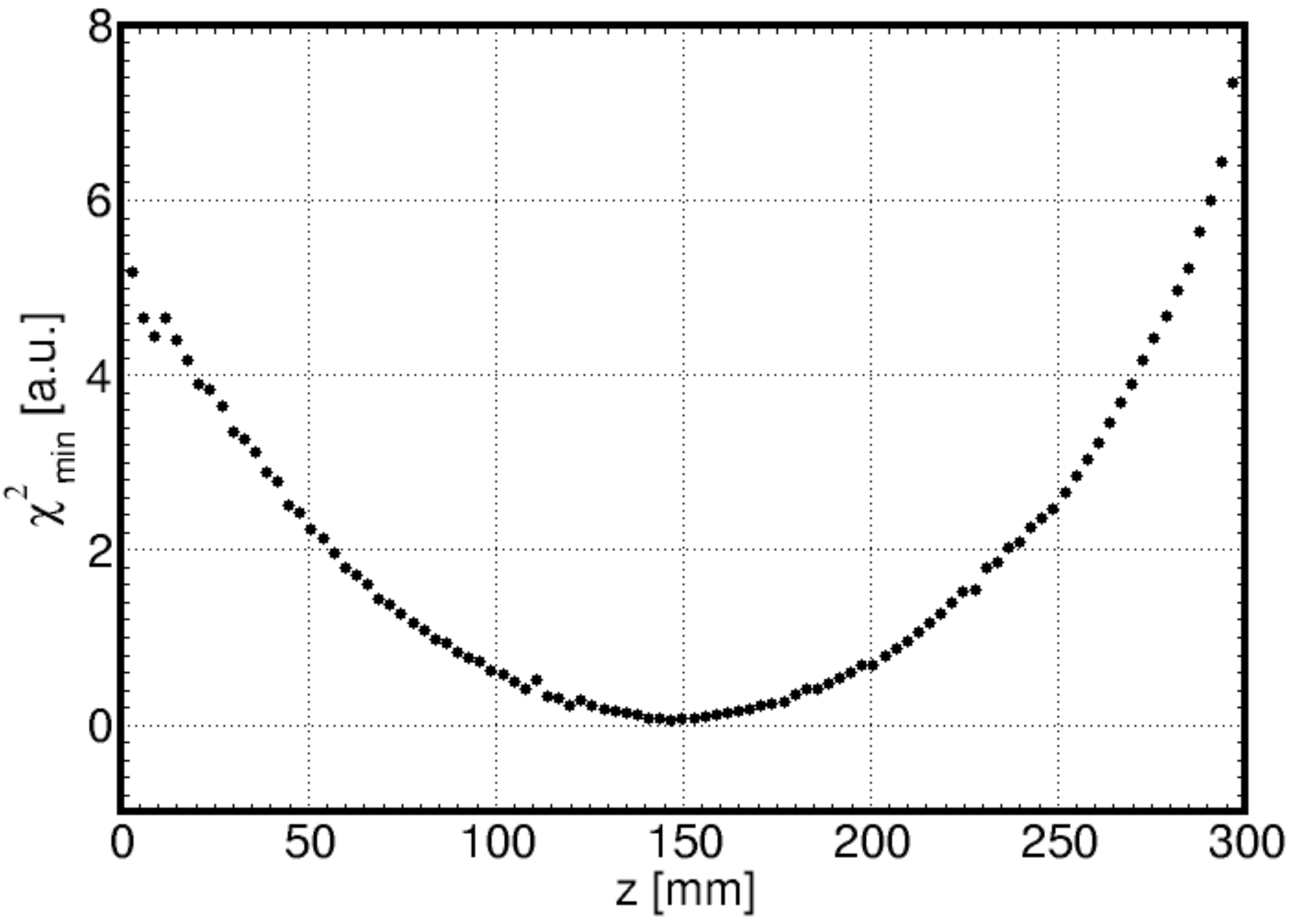}
 \end{center}
 \caption{Example of the $\chi^2$ distribution defined in Eq.~\ref{eqn3}}
\label{Fig6}
 \end{figure}
An example of the $\chi^2$ distribution calculated according to Eq.~\ref{eqn3} for one of signals sampled with constant
levels and measured at central position of the strip is shown in Fig.~\ref{Fig6}. One can see a clear minimum corresponding
to z~$\approx$~150~mm.
\begin{figure*}[h]

           \centerline{
           \subfigure[!htb][]
           {
              \includegraphics*[width=0.5\textwidth]{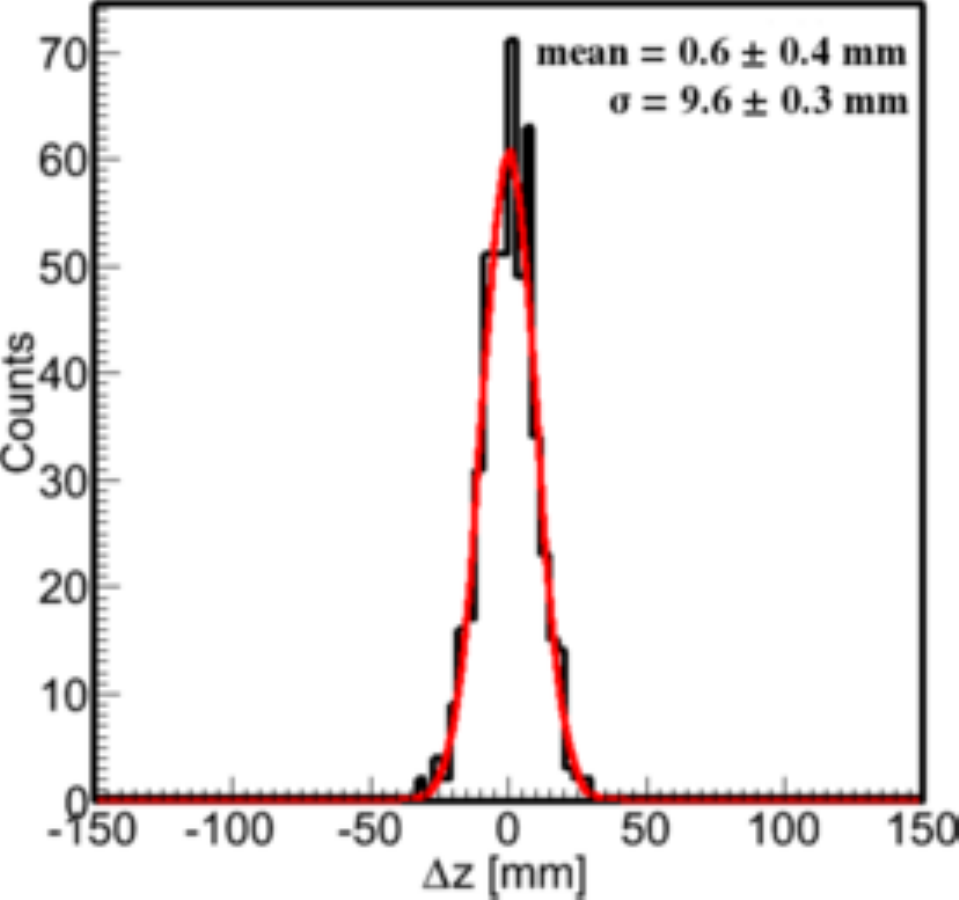}
               \label{Figg(a)}
           }
           \subfigure[!htb][]
           {
               \includegraphics*[width=0.5\textwidth]{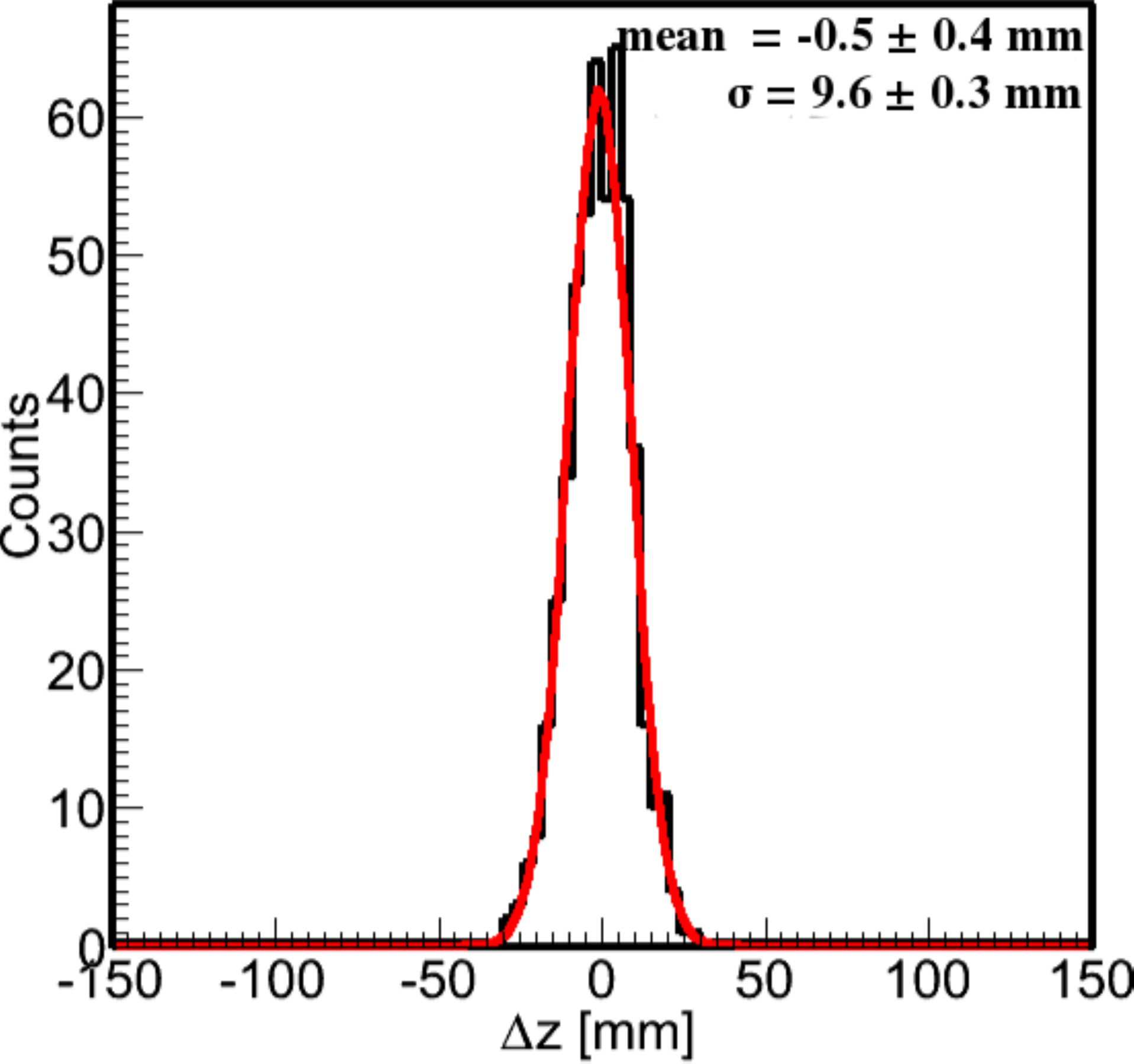}}
                \label{Figg(b)}
           }
              \caption{Distribution of differences between the true and reconstructed position $\Delta z$ for
signals  measured at z = 150 mm for (a) constant fractions (b) constant levels.} \label{Fig7}
\end{figure*}
Present version of reconstruction procedure do not take into account the measurement uncertainties. Therefore, the $\chi^2$
values are treated as arbitrary and errors of fitted parameters are determined directly from distributions of differences
between reconstructed and true values of time or position. Figure.~\ref{Fig7}a and b show distributions of differences
between the true and reconstructed position using $\chi^2$ defined by Eq.~\ref{eqn3} for constant fraction and constant
levels discrimination, respectively.
%Gaussian fits to these distribution give the
Resolution of position reconstruction is determined by fitting a gauss function to the presented distribution and obtained
results give: $\sigma_{z} \approx$~9.7$\pm$ 0.3 mm for constant fraction sampling and $\sigma_{z} \approx$~9.6 $\pm$ 0.3 mm
for constant levels.
%%%%%%%%%%%%%%%%%%%%%%%%%%%%%%%%%%%%%
\clearpage
%%%%%%%%%%%%%%%%%%%%%%%%%%%%%%%%%%%%%
%
\begin{figure*}[h]
           \centerline{
           \subfigure[!htb][]
           {
              \includegraphics*[width=0.5\textwidth]{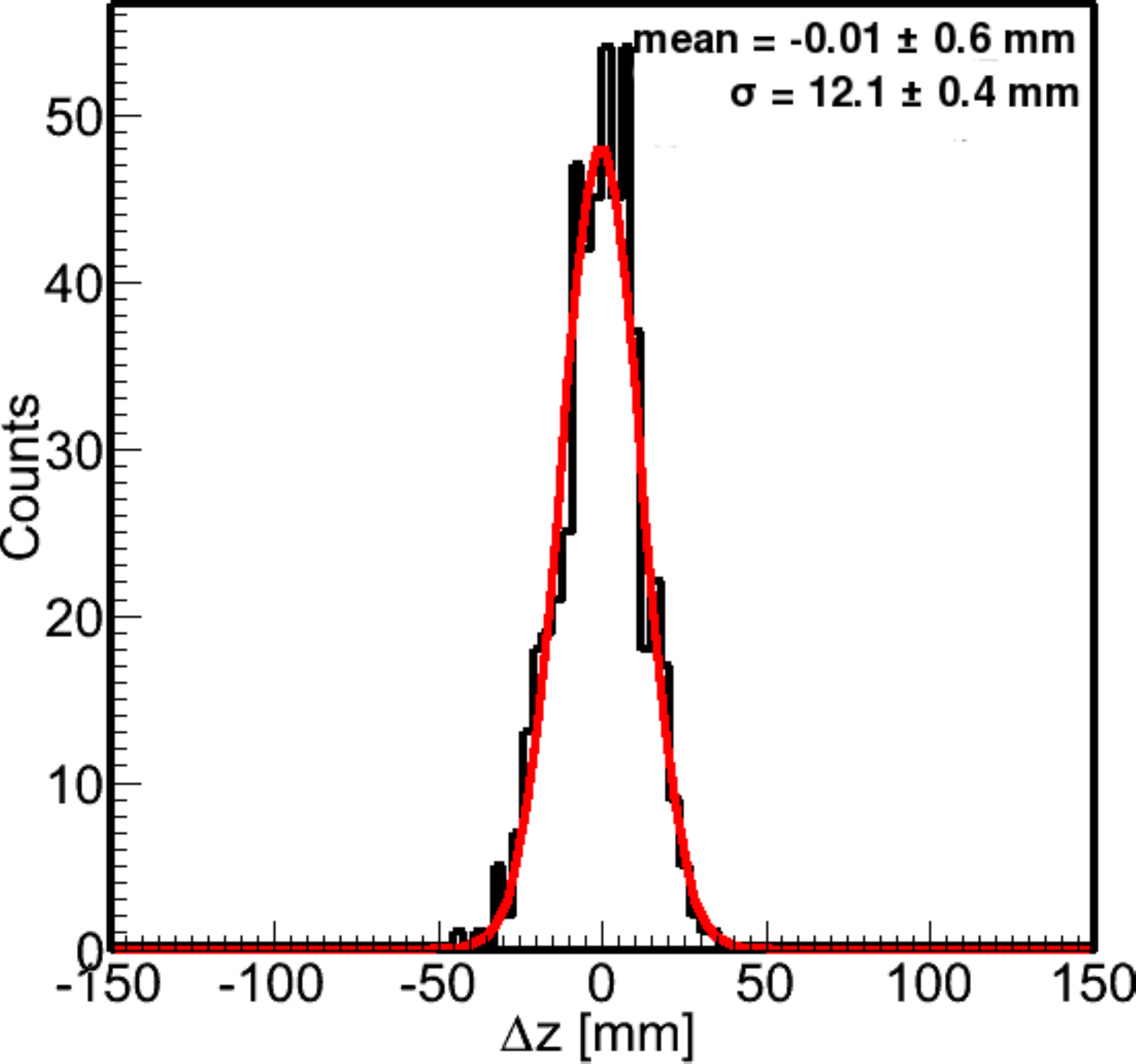}
               \label{Fiig(a)}
           }
           \subfigure[!htb][]
           {
               \includegraphics*[width=0.5\textwidth]{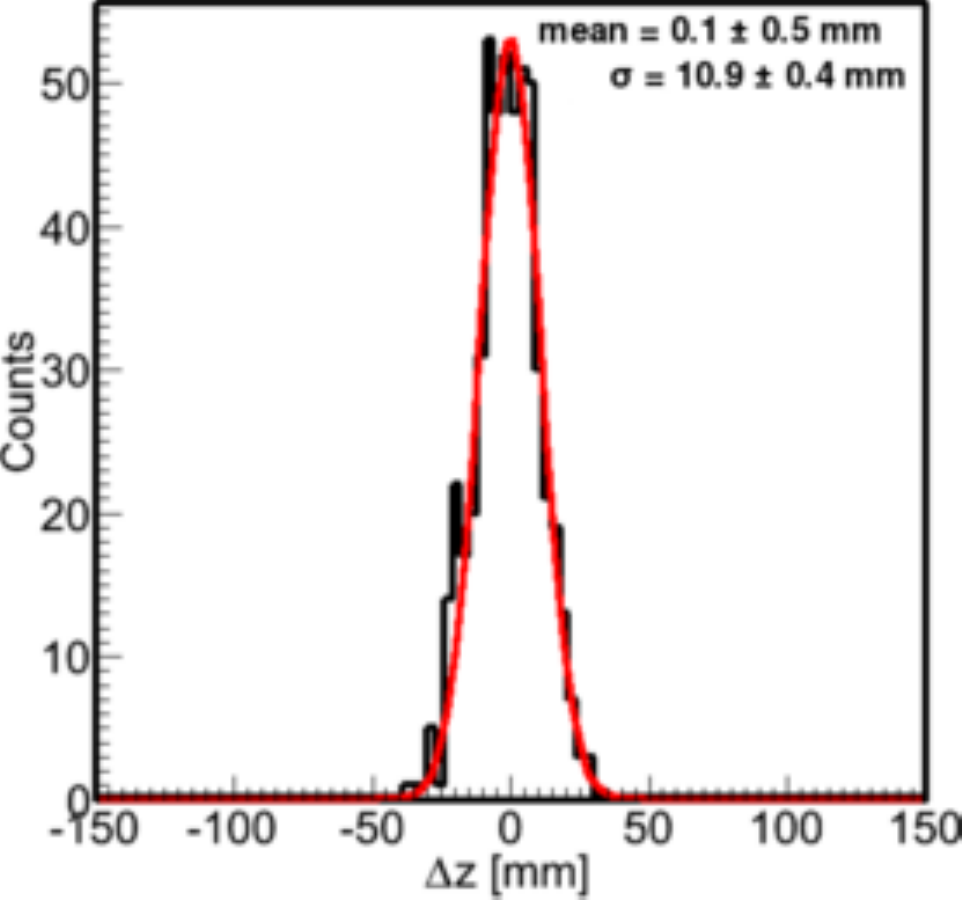}}
                \label{Fiig(b)}
           }
              \caption{Distribution of differences between the true and reconstructed position $\Delta z$
              for signals measured at z = 150 mm for (a) constant fractions (b) constant levels}
\label{Fig8}
\end{figure*}
 In case when $\chi^2$ is a function of $\delta$t, $\alpha_L$ and $\alpha_R$ (Eq.~\ref{eqn3}) obtained
position resolution amounts to $\sigma_{z} \approx$~12.1 $\pm$ 0.4 mm and 10.9 $\pm$ 0.4 mm for constant fraction and
constant level method, respectively. Corresponding $\Delta z$ distributions are shown in Fig.~\ref{Fig8}.
 \begin{figure*}[h]
           \centerline{
           \subfigure[!htb][]
           {
              \includegraphics*[width=0.5\textwidth]{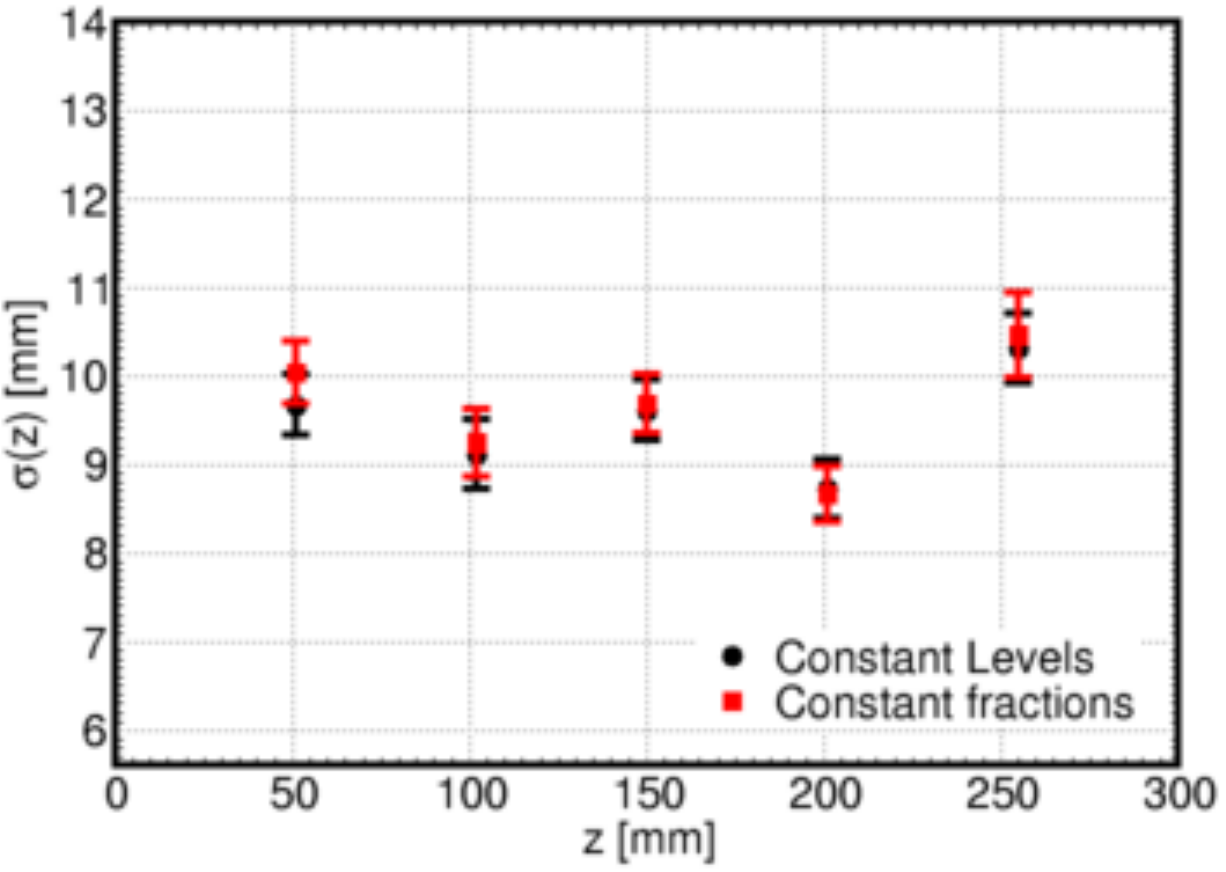}
               \label{Ffig(a)}
           }
           \subfigure[!htb][]
           {
               \includegraphics*[width=0.5\textwidth]{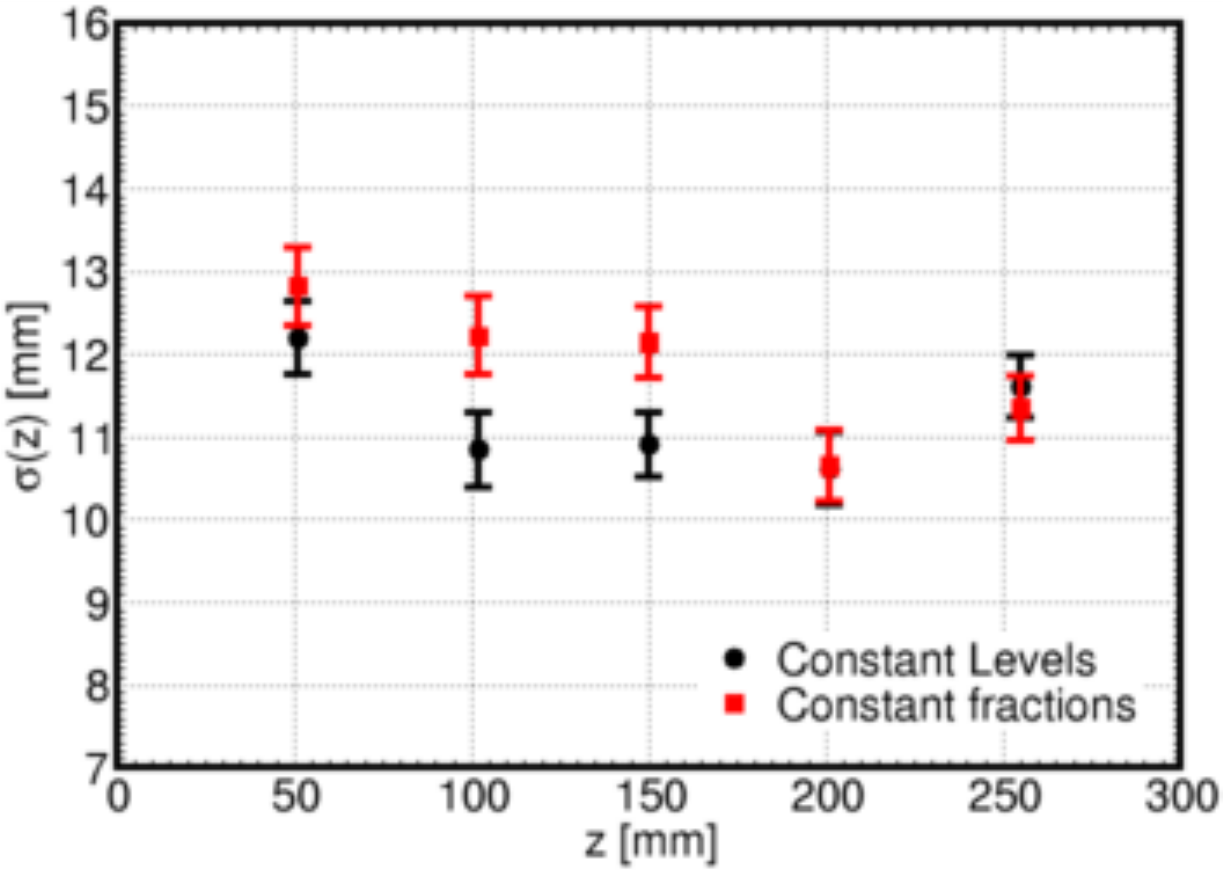}}
                \label{Ffig(b)}
           }
              \caption{Position resolution as a function of the position of gamma quantum interaction for the case when
 (a) $\chi^2$ is a function of $\delta$t only (b) $\chi^2$ is a function of $\delta$t,
              $\alpha_L$ and $\alpha_R$.}
\label{Fig9}
\end{figure*}
These resolutions were determined for signals measured at several positions along the scintillator as it is shown in
Fig.~\ref{Fig9}. The results indicate that this resolution does not change much with position.
Results for TOF reconstruction for signals measured at z~=~150~mm are shown in Fig.~\ref{Fig10}. Corresponding resolutions
are in this case equal to $\sigma_{TOF}\sim$~163~ps for constant fractions sampling and $\sigma_{TOF}\sim$~143~ps in case
of constant levels.
%%%%%
\begin{figure*}[h]
           \centerline{
           \subfigure[!htb][]
           {
              \includegraphics*[width=0.5\textwidth]{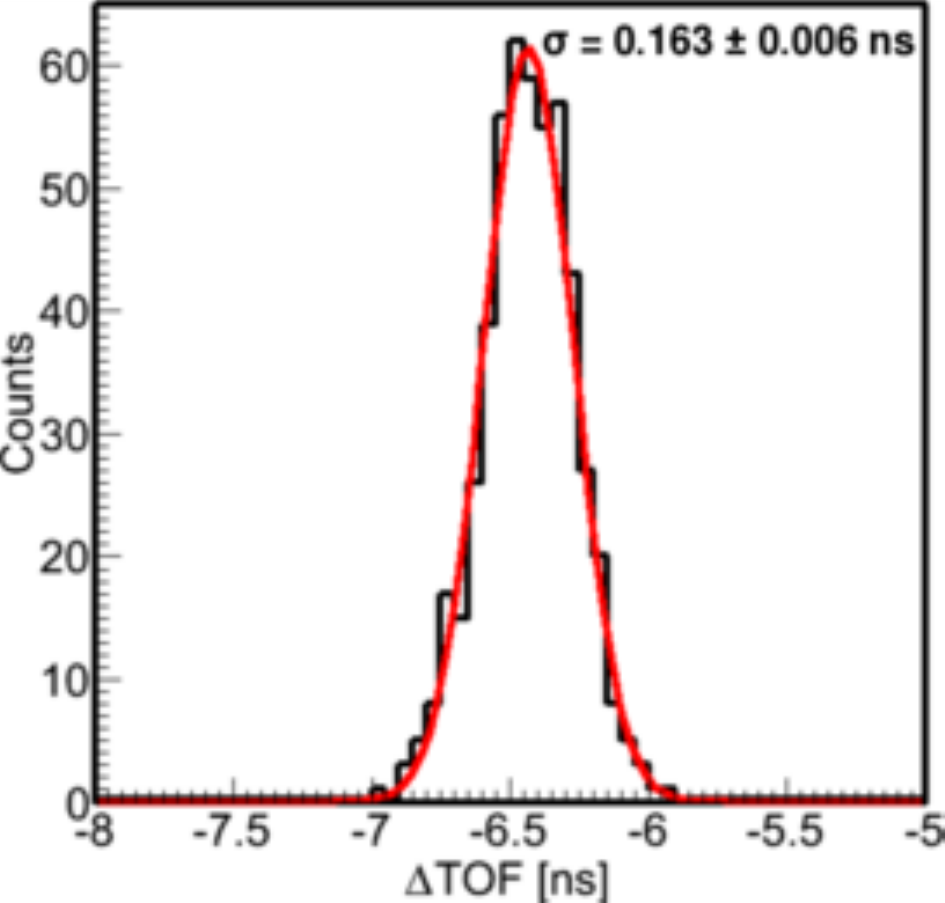}
               \label{Figgg(a)}
           }
           \subfigure[!htb][]
           {
               \includegraphics*[width=0.5\textwidth]{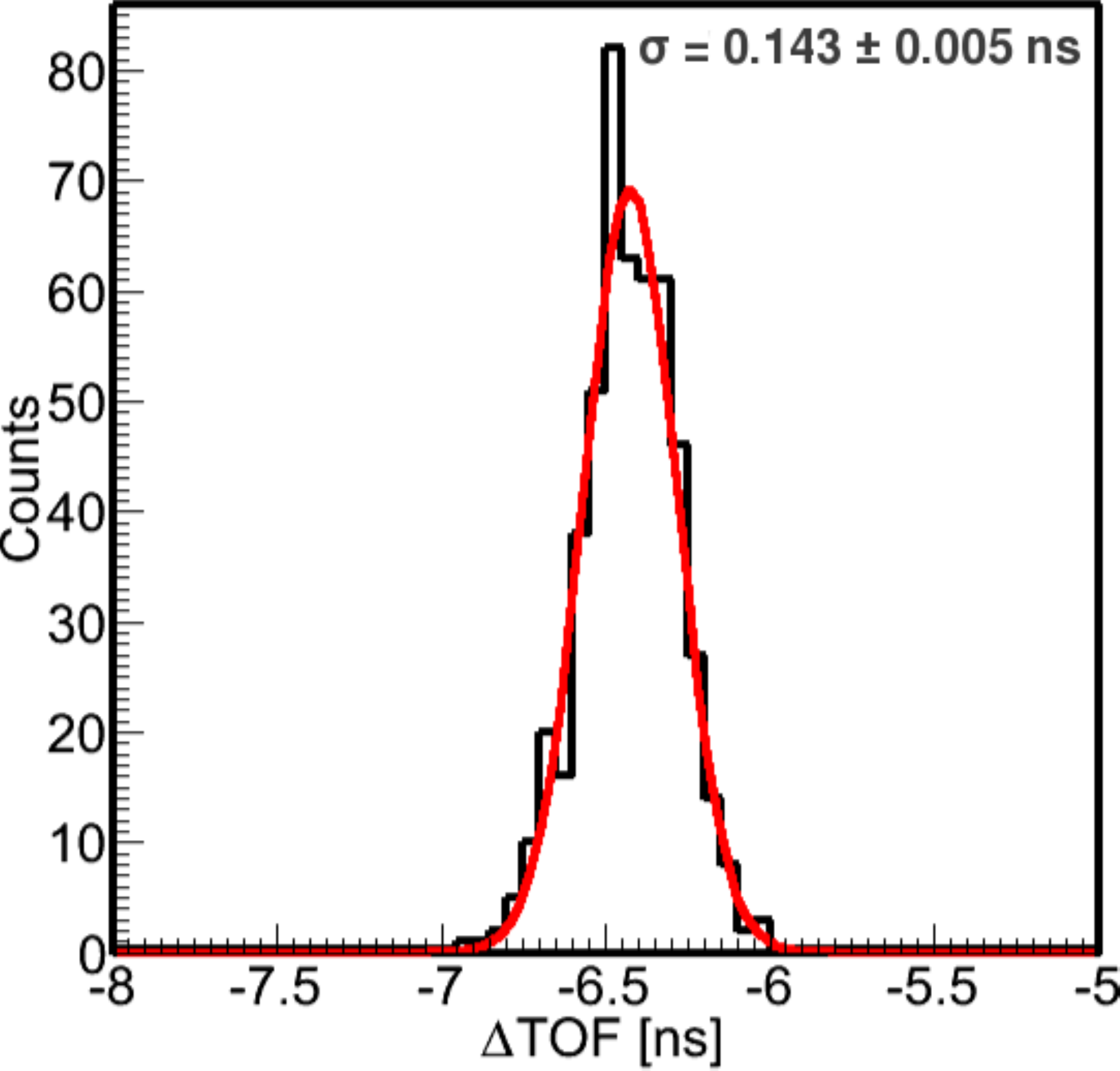}}
                \label{Figgg(b)}
           }
              \caption{Distribution of differences between the true and reconstructed TOF
                       for events registered at z~=~150~mm. The reconstruction was done using $\chi^2$
                       as a function of $\delta t$ with signal sampling at (a) constant fraction and
                       (b) at constant levels}
\label{Fig10}
\end{figure*}
As it is shown in Fig.~\ref{Fig11} results obtained for $\chi^2(\delta t, \alpha_L$, $\alpha_R)$ amount to
$\sigma_{TOF}\sim$~132~ps (constant fractions) and $\sigma_{TOF}\sim$~119~ps (constant levels).
\begin{figure*}[b]
           \centerline{
           \subfigure[!htb][]
           {
              \includegraphics*[width=0.5\textwidth]{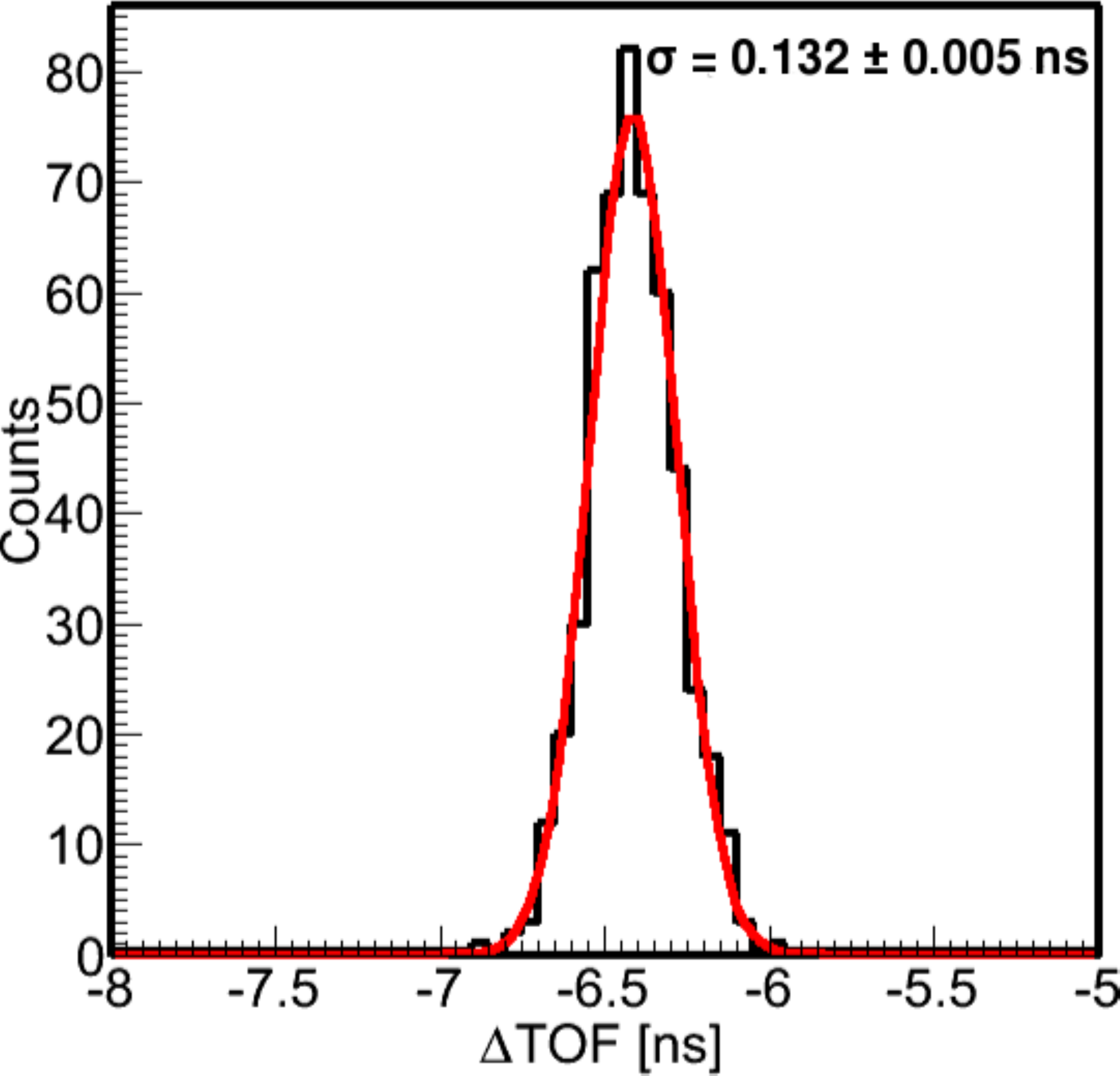}
               \label{Fiiig(a)}
           }
           \subfigure[!htb][]
           {
               \includegraphics*[width=0.5\textwidth]{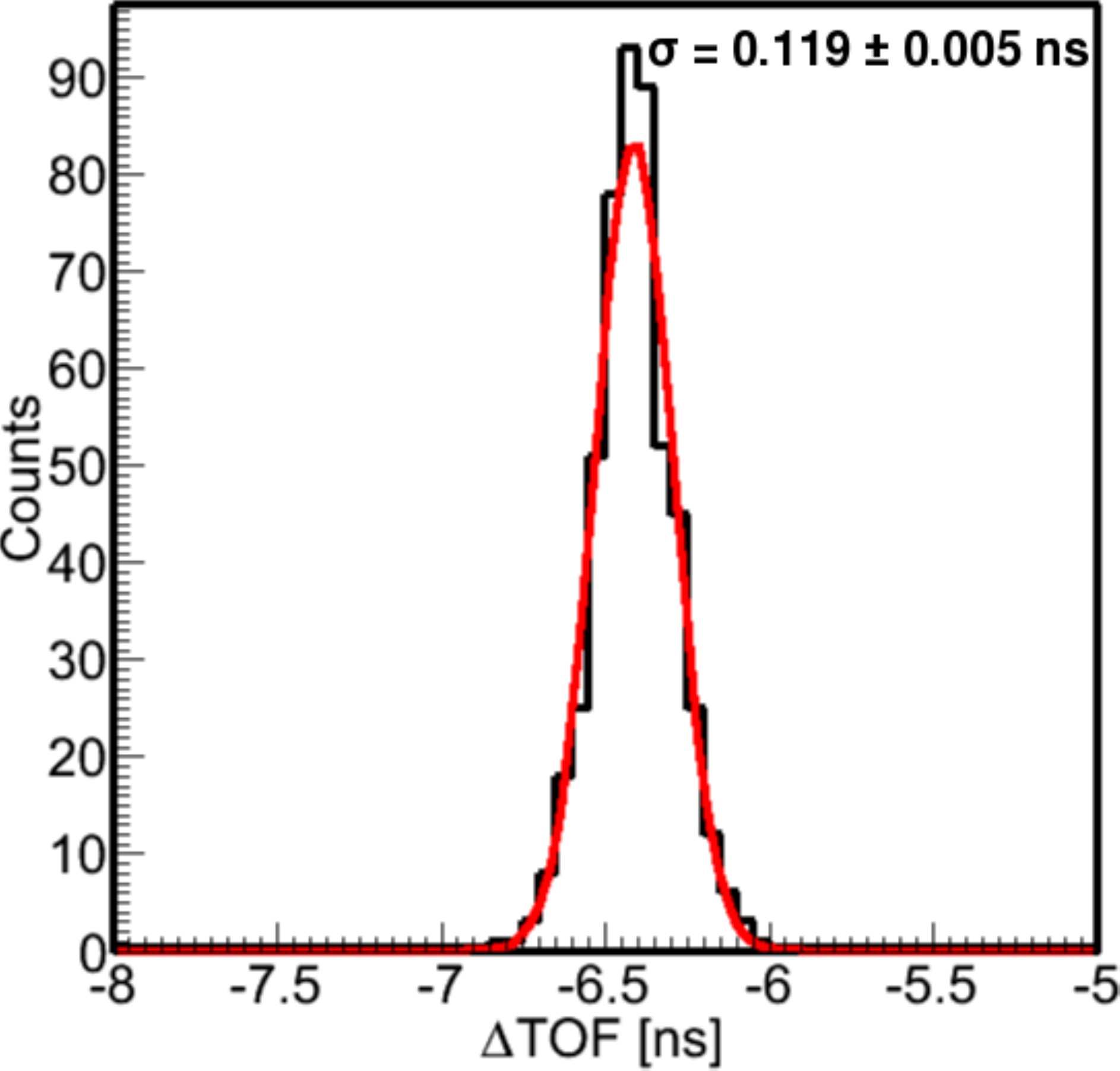}}
                \label{Fiiig(b)}
           }
              \caption{Distribution of differences between the true and reconstructed TOF for
                       signals measured at z = 150 mm. The reconstruction was done using $\chi^2$ as a function of $\delta$t,$\alpha_L$, and
                       $\alpha_R$, and with signal sampling at(a) constant fraction(b) constant levels.} \label{Fig11}
\end{figure*}
%%%%%%%%%%%%%%%%%%%%%%%%%%%%%%%%
\clearpage
%%%%%%%%%%%%%%%%%%%%%%%%%%%%%%%
In principle true value of TOF should be equal to zero when source was positioned in the middle between
detection modules. However, due to different time offsets produced by electronics the reconstructed mean values of TOF may
be different from zero.
\begin{figure*}[h]
           \centerline{
           \subfigure[!htb][]
           {
              \includegraphics*[width=0.5\textwidth]{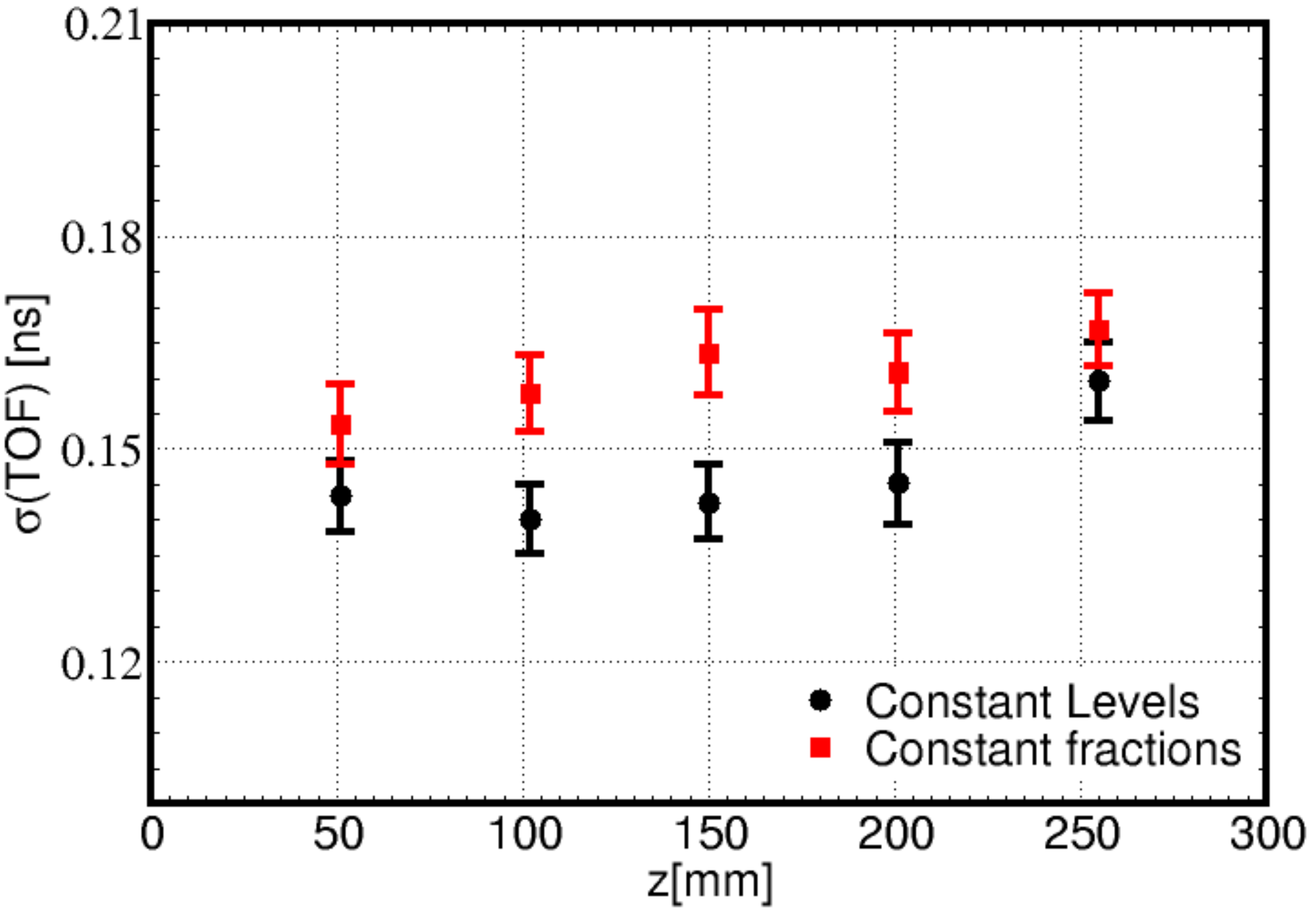}
               \label{Fffig(a)}
           }
           \subfigure[!htb][]
           {
               \includegraphics*[width=0.5\textwidth]{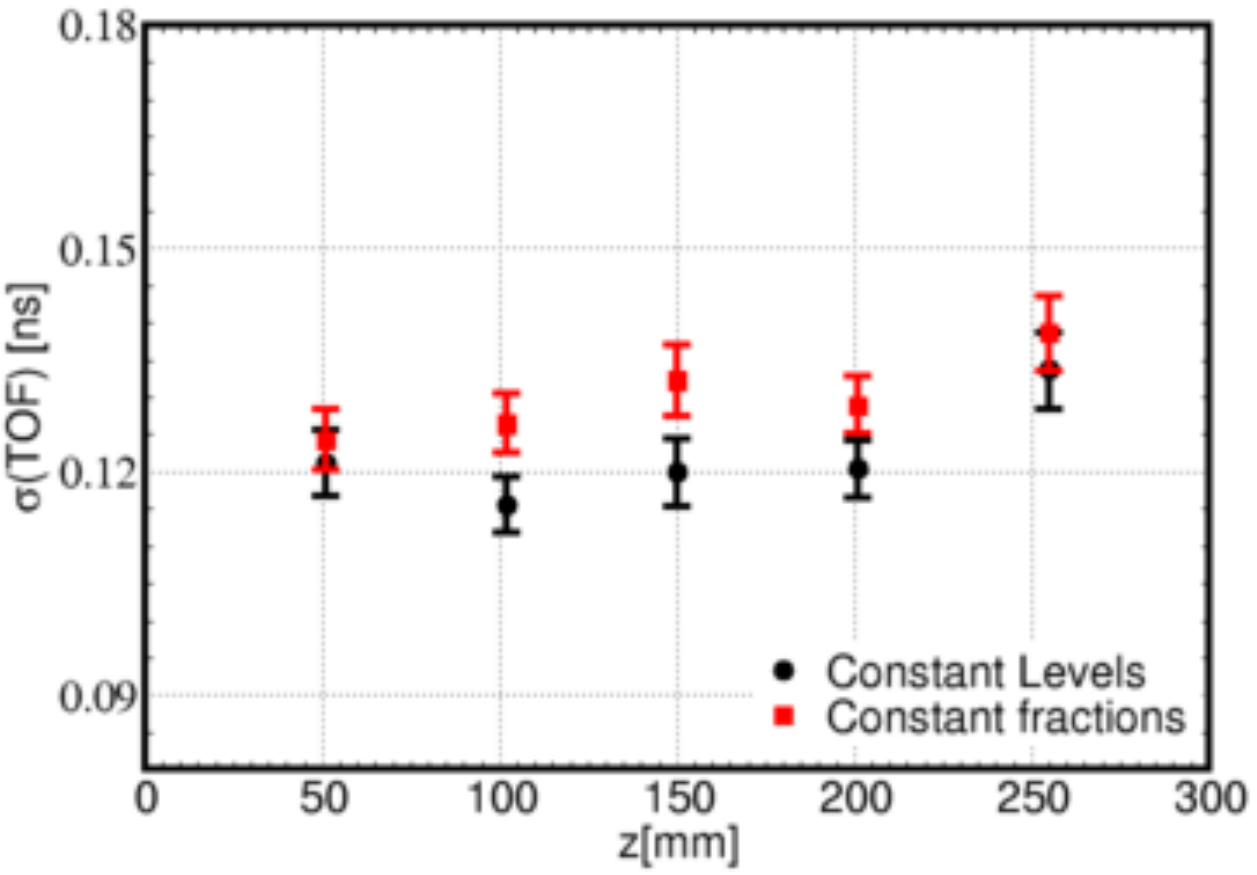}}
                \label{Fffig(b)}
           }
              \caption{TOF resolution as a function of position along the scintillator:(a) $\chi^2 = \chi^2(\delta t)$ and (b)  $\chi^2 = \chi^2(\delta t, \alpha_L, \alpha_R)$.} \label{Fig12}
\end{figure*}
Again, we have performed studies of the TOF resolution as a function of position for both $\chi^2$ functions as it is shown
in Fig~\ref{Fig12}. One can see that the determined resolutions is constant within the error bars over the full length of
30~cm long scintillator strip.
\section{Summary}
\label{sum} The preliminary results obtained during validation of the reconstruction method introduced in this article show
that it is possible to obtain a spatial resolution of about 1.2~cm ($\sigma$) for the gamma quanta hit position, and TOF
resolution of about 125~ps ($\sigma$). As regards the position resolution along the detector obtained result is few time
worse than achieved by the commercial TOF-PET scanners, However as regards the TOF determination obtained result is better
by about a factor of two with respect to resolutions achieved by the commercial TOF-PET tomographs characterized by typical
field of views of about 16~cm and TOF resolution of about 230 ps ($\sigma$)~\cite{V}. A further improvement is expected in
the future by including measurement uncertainties and possible correlations between the times measured at different
thresholds.

\end{document}